\documentclass[12pt]{article}
\usepackage{graphicx,amssymb,amsmath}

\newcommand{\be}{\begin{equation}}
\newcommand{\ee}{\end{equation}}
\newcommand{\bea}{\begin{eqnarray}}
\newcommand{\eea}{\end{eqnarray}}
\newcommand{\beas}{\begin{eqnarray*}}
\newcommand{\eeas}{\end{eqnarray*}}

\newcommand{\x}{{\bf x}}

\begin{document}
\begin{titlepage}

\begin{center}

{\Large CFT representation of interacting bulk gauge fields in AdS}

\vspace{8mm}

\renewcommand\thefootnote{\mbox{$\fnsymbol{footnote}$}}
Daniel Kabat${}^{1}$\footnote{daniel.kabat@lehman.cuny.edu},
Gilad Lifschytz${}^{2}$\footnote{giladl@research.haifa.ac.il}

\vspace{4mm}

${}^1${\small \sl Department of Physics and Astronomy} \\
{\small \sl Lehman College, City University of New York, Bronx NY 10468, USA}

\vspace{2mm}

${}^2${\small \sl Department of Mathematics and Physics} \\
{\small \sl University of Haifa at Oranim, Kiryat Tivon 36006, Israel}

\end{center}

\vspace{8mm}

\noindent
We develop the representation of interacting bulk gauge fields and
charged scalar matter in AdS in terms of non-local observables in the
dual CFT.  We work in holographic gauge in the bulk, $A_z = 0$.  The
correct statement of micro-causality in holographic gauge is somewhat
subtle, so we first discuss it from the bulk point of view.  We then
show that in the $1/N$ expansion CFT correlators can be lifted to
obtain bulk correlation functions which satisfy micro-causality.  This
requires adding an infinite tower of higher-dimension multi-trace
operators to the CFT definition of a bulk observable.  For conserved
currents the Ward identities in the CFT prevent the construction of
truly local bulk operators (i.e.\ operators that commute at spacelike
separation with everything), however the resulting non-local
commutators are exactly those required by the bulk Gauss constraint.
In contrast a CFT which only has non-conserved currents can be lifted
to a bulk theory which is truly local.  Although our explicit
calculations are for gauge theory, similar statements should hold for
gravity.

\end{titlepage}
\setcounter{footnote}{0}
\renewcommand\thefootnote{\mbox{\arabic{footnote}}}

\section{Introduction\label{sect:intro}}

The question of observables in quantum gravity has a long history; for
reviews see
\cite{Rovelli:1990ph,Kuchar:1991qf,Isham:1992ms,Tambornino:2011vg}.
The problem is that, as emphasized by Dirac \cite{Dirac}, only
gauge-invariant quantities can be assigned a physical meaning.  In
gravity this rules out the existence of local observables.  Indeed in
the AdS/CFT context a complete set of observables lives at the
boundary, so one must be able to express any definition of a bulk
observable in terms of CFT data.  In the limit of free scalar fields
in the bulk (i.e.\ zero Planck length, $N \rightarrow \infty$) the
construction was made in the early days of AdS/CFT \cite{Banks:1998dd,
Balasubramanian:1998sn, Dobrev:1998md, Bena:1999jv}. It has been
recast in the form of a smearing function \cite{Hamilton:2005ju,
Hamilton:2006az}
\begin{equation}
\phi(z,x)=\int dx' K_{\Delta}(z,x|x'){\cal O}_{\Delta}(x')
\label{basic}
\end{equation}
where the kernel $K$ has support only on boundary points $x'$ which
are space-like separated from the bulk point $(z,x)$.  The dimension
of the boundary operator $\Delta$ is determined by the mass of the
bulk field. It turns out that using complex boundary coordinates is a
very convenient computational tool \cite{Hamilton:2006az,
Hamilton:2006fh}. These constructions were carried out in the free
field limit, and it was shown that the CFT expectation value of two
such operators reproduces the free bulk two point function.

Building on these works, the construction of interacting bulk
observables in terms of smeared CFT operators has been developed.  Two
approaches have been worked out, both relying on perturbation theory
in $1/N$. One approach is based on the bulk equations of motion, while
the other uses bulk micro-causality as a guiding principle.

The first approach was introduced in \cite{Kabat:2011rz} and
further developed in \cite{Heemskerk:2012mn}.  The basic idea is to
solve the bulk equations of motion perturbatively.  This can be done
in a fixed gauge (holographic gauge), using the radial supergravity
Hamiltonian on a fixed background.  This procedure gives a bulk
operator written in terms of smeared CFT operators, whose correlation
functions in the CFT reproduce bulk correlators.  This construction
can be carried out independently of holography. It is just a rewriting
of bulk correlation functions in terms of boundary correlators, in the
same way that one could have computed a bulk correlator in terms of
correlation functions on some initial time slice by solving time
evolution equations.  The only difference is that in AdS/CFT it is
convenient to evolve in a spacelike direction, using a spacelike
Green's function \cite{Kabat:2011rz, Heemskerk:2012mn}. It is an extra
condition, that the boundary correlation functions are those of a
unitary CFT, which makes the relationship holographic.  But in the
approach of solving bulk equations of motion, the role played by
holography is not so clear.

In the second approach, more intrinsic to the CFT, one tries to build
up the bulk operator by requiring that it satisfy bulk micro-causality
\cite{Kabat:2011rz}.\footnote{For a discussion of micro-causality in
curved space see \cite{Dubovsky:2007ac}.}  This program was carried
out for scalar fields in \cite{Kabat:2011rz}, where the requirement of
micro-causality is just that bulk operators commute at space-like
separation. The basic point is that, if one inserts the smeared CFT
operator (\ref{basic}) inside a CFT three point function, there are in
general singularities at bulk space like separation.  These
singularities lead to a non-zero commutator which spoils
micro-causality. However these singularities can be suppressed (in a
precise sense) by redefining the bulk operator to include an infinite
tower of appropriately smeared higher dimension scalar primaries
(these are the multi-trace operators also discussed in
\cite{Heemskerk:2009pn}),
\begin{equation}
\phi(z,x)=\int dx' K_{\Delta}(z,x|x'){\cal O}_{\Delta}(x')+\sum_{l} a_{l} \int dx' K_{\Delta_{l}}(z,x|x'){\cal O}_{\Delta_{l}}(x').
\end{equation}
Order-by-order in $1/N$ one has the required spectrum of higher
dimension operators, and one can choose the coefficients $a_{l}$ in
such a way that the bulk operator satisfies micro-causality.  The
resulting bulk observable agrees with what one would construct in the
$1/N$ expansion by solving the bulk equations of motion perturbatively.

The CFT approach gives a different perspective from the approach based
on bulk equations of motion, and gives a glimpse of the non-locality
which is expected when the boundary theory is a finite-$N$ unitary
CFT. More specifically, the CFT construction requires the existence of
an infinite tower of higher-dimension primary operators with
prescribed properties.  Such operators can be constructed in $1/N$
perturbation theory as multi-trace operators, but these operators do
not actually exist in a unitary CFT at finite $N$. Thus at finite $N$
we can see how bulk locality breaks down in a non-perturbative way.

In extending the CFT construction of interacting bulk observables to
include gravity we face the difficulty mentioned at the start of the
introduction, that in gravity there are no local physical observables
(see \cite {Heemskerk:2012mn,Heemskerk:2012mq,Kabat:2012hp} for
discussions of this in the context of AdS/CFT).  There is, however,
clearly some sense in which local observables are available even in a
theory of gravity.  To address this we need to deal
with the underlying gauge symmetry.  There are two approaches we could
take: either construct a set of gauge-invariant observables, or carry
out the construction in a fixed gauge.  We will adopt the gauge-fixed
approach, which may not seem so elegant but is in fact natural in
AdS/CFT.

Of course the two approaches are related.  To be concrete, consider a
charged scalar field in the bulk $\phi(x,z)$, coupled to an abelian
gauge field $A$.  Here $x=(t,\vec{x})$ are coordinates in the CFT and
$z$ is a radial coordinate.  We work in holographic gauge which sets
$A_z = 0$.  In holographic gauge $\phi({x},z)$ is (by definition) a
gauge-invariant observable.  It can be identified with the manifestly
gauge-invariant quantity
\[
\exp \Big[{i \int_{({ x},z)}^{({x},0)} A_z dz}\Big] \phi({x},z)
\]
where we have attached a Wilson line running from the bulk point to
the boundary of AdS in the $z$ direction.  But now one sees the
difficulty, that although a bulk scalar field in holographic gauge is
an observable quantity, it is secretly non-local, with a Wilson line
extending in the $z$ direction.  So there is no reason to expect our
gauge-fixed operators to commute at spacelike separation, and indeed
in axial gauge there are non-local commutators \cite{Hanson:1976cn}.

In gauge theory it's tempting to avoid this issue by working in terms
of local gauge-invariant quantities, such as ${\rm Tr} \, F^2$ or
$\phi^\dagger \phi$, but in gravity this is not an option.  So let's
work directly with the scalar field in holographic gauge, and see if
there is a useful sense in which we can discuss bulk
locality.\footnote{An alternative approach would be to work in some
type of covariant gauge in the bulk, where locality is manifest but
there are additional unphysical degrees of freedom.  It's not clear
to us how this could be represented in the CFT.}

A key observation is that in holographic gauge non-local commutators are
indeed present, but {\em only to the extent required by the
constraints}.  For example consider a charged scalar field $\phi({
x},z)$ and the electric flux observable
\[
\Phi_E = \oint {}^*F
\]
Since charge can be measured by a surface integral arbitrarily far
away, it's clear that $\phi({ x},z)$ and $\Phi_E$ will in general not
commute at equal times.  But there's no obstacle to having $\phi$
commute with itself at equal times.  We will make this more precise in
section \ref{sect:bulk}, where we consider scalar electrodynamics in
holographic gauge and show that the scalar field indeed commutes with
itself at spacelike separation.  There are some non-local commutators
in holographic gauge.  However at equal times the only non-local
commutators involve either the time component of the gauge field $A_0$
or the $z$ component of the electric field $E_z$.  This behavior is
exactly what's required by the Gauss constraint, and it can be
understood as due to the Wilson lines extending in the $z$
direction.\footnote{The extension to gravity seems clear: scalar
fields will commute with each other at spacelike separation, but
they will have non-zero commutators with $h_{00}$ and with certain
components of the curvature.}

Our conclusion is that we can construct bulk observables in
holographic gauge by demanding that, for example, charged bulk scalar fields
commute at spacelike separation.  Of course gauge-invariant
combinations such as a field strength in the bulk and a scalar field
on the boundary will also commute at spacelike separation.  The
remainder of this paper is devoted to showing that these requirements,
which we view as encoding bulk micro-causality, suffice to uniquely
determine the way in which boundary CFT correlators can be lifted into
the bulk.

\section{Bulk micro-causality\label{sect:bulk}}

In this section we consider scalar electrodynamics in holographic
gauge and show that the scalar field commutes with itself at spacelike
separation.  Our treatment of the canonical formalism for scalar
electrodynamics in holographic gauge closely follows section 5.C of
\cite{Hanson:1976cn}.

We work in AdS${}_{d+1}$ with metric 
\[
ds^2 = {R^2 \over z^2} \left(-dt^2 + \vert d\vec{x} \vert^2 + dz^2\right)
\]
and consider scalar electrodynamics with action ($D_M = \partial_M + i q A_M$)
\be
S = \int d^{d+1}x \sqrt{-g} \left(-D_M \phi^* D^M \phi - {1 \over 4} F_{MN} F^{MN}\right)
\label{ScalarElectrodynamics}
\ee
The canonical momenta are
\beas
&& \pi_0 = 0 \\
&& \pi_i = \left({R \over z}\right)^{d-3} \left(\partial_0 A_i - \partial_i A_0\right) \qquad i = 1,\ldots,d \\
&& \pi_\phi = \left({R \over z}\right)^{d-1} D_0 \phi^* \\
&& \pi_\phi^* = \left({R \over z}\right)^{d-1} D_0 \phi
\eeas
Thus we have the primary constraint
\[
\chi_1 \equiv \pi_0 = 0
\]
and the secondary constraint (Gauss' law)
\[
\chi_2 \equiv \partial_i \pi_i + iq \left(\pi_\phi \phi - \pi_\phi^* \phi^*\right) = 0
\]
Conjugate to these we impose the two gauge-fixing conditions
\beas
&& \chi_3 \equiv A_z = 0 \\
&& \chi_4 \equiv \pi_z + \left({R \over z}\right)^{d-3} \partial_z A_0 = 0
\eeas
The first condition fixes holographic gauge, while the second condition enforces the usual relation between the
$z$ component of the electric field and the gauge field.

The matrix of Poisson brackets is (setting $\x = ({ \vec{x}},z)$)
\[
\hspace{-1.7cm} C_{ab} = \lbrace \chi_a(\x), \chi_b(\x') \rbrace = \left(
\begin{array}{cccc}
0 & 0 & 0 & -({R \over z'})^{d-3} \partial_z \delta^d(\x - \x') \\
0 & 0 & \partial_z \delta^d(\x - \x') & 0 \\
0 & \partial_z \delta^d(\x - \x') & 0 & - \delta^d(\x - \x') \\
-({R \over z'})^{d-3} \partial_z \delta^d(\x - \x') & 0 & \delta^d(\x - \x') &  0
\end{array}\right)
\]
This has an inverse
\be
C^{-1}_{ab} = \left(\begin{array}{cccc}
0 & g(\x,\x') & 0 & ({z' \over R})^{d-3} f(\x,\x') \\
-g(\x,\x') & 0 & -f(\x,\x') & 0 \\
0 & -f(\x,\x') & 0 & 0 \\
({z \over R})^{d-3} f(\x,\x') & 0 & 0 & 0
\end{array}\right)
\ee
where
\bea
\label{fg}
&& f(\x,\x') = \delta^{d-1}({ x} - { x}') \theta(z'-z) \\
\nonumber
&& g(\x,\x') = \delta^{d-1}({ x} - { x}') \theta(z'-z) {(z')^{d-2} - z^{d-2} \over (d-2) R^{d-3}}
\eea
The inverse is not unique; our explicit choice for $f$ and $g$
corresponds to introducing a Wilson line towards the boundary of AdS,
as opposed to towards the Poincar\'e horizon.  Also note that the case
$d = 2$ is special as it corresponds to Chern-Simons theory in the
bulk \cite{Jensen:2010em}.  For the canonical formalism in this case
see appendix B of \cite{Kabat:2012hp}.

Given the structure of the constraint algebra -- in particular the fact that $C^{-1}_{22} = 0$ -- it follows that the physical
degrees of freedom have canonical Dirac brackets at equal times.
\beas
&& \lbrace \pi_{\hat{\imath}}(\x), A_{\hat{\jmath}}(\x') \rbrace = \delta_{\hat{\imath}\hat{\jmath}} \, \delta^d(\x - \x') \qquad \hat{\imath},\hat{\jmath} = 1,\ldots,d-1 \\
&& \lbrace \pi_\phi(\x), \phi(\x') \rbrace = \delta^d(\x-\x') \\
&& \lbrace \pi_\phi^*(\x), \phi^*(\x') \rbrace = \delta^d(\x-\x')
\eeas
However these fields have non-local Dirac brackets with $A_0$ and $\pi_z$, namely
\beas
&& \lbrace A_0(\x), A_{\hat{\imath}}(\x') \rbrace = \partial_{\hat{\imath}} g(\x,\x') \\
&& \lbrace A_0(\x), \phi(\x') \rbrace = i q g(\x,\x') \phi(\x') \\
&& \lbrace A_0(\x), \pi_\phi(\x') \rbrace = - i q g(\x,\x') \pi_\phi(\x') \\[10pt]
&& \lbrace \pi_z(\x), A_{\hat{\imath}}(\x') \rbrace = \partial_{\hat{\imath}} f(\x,\x') \\
&& \lbrace \pi_z(\x), \phi(\x') \rbrace = i q f(\x,\x') \phi(\x') \\
&& \lbrace \pi_z(\x), \pi_\phi(\x') \rbrace = - i q f(\x,\x') \pi_\phi(\x')
\eeas
along with the complex conjugates.  These brackets reflect the fact
that the field $\phi({ x},z)$ produces a tube of electric flux
extending towards $z = 0$.

This shows that, as promised, the scalar field commutes with itself at
equal times.  However we'd like to make a stronger statement, that the
scalar field commutes with itself at spacelike separation.  This can be
argued based on the results obtained above.  Imagine inserting the
scalar field at two spacelike separated points $(x,z)$ and $(x',z')$,
with Wilson lines secretly extending off in the $z$ direction.  By
acting with an AdS isometry the two bulk points can be brought to
equal times.  However the isometry will act on the Wilson lines, so
they will no longer extend in the $z$ direction.  We could perform a
compensating gauge transformation to restore holographic gauge $A_z =
0$, but it's simpler to leave the Wilson lines pointing in whatever
direction is implied by the isometry.  How would this affect the above
calculation?  The brackets with $A_0$ and $\pi_i$ will clearly be
different, because the electric flux tubes now go in a different
direction, but the bracket of $\phi$ with itself will still be zero.
This means that in holographic gauge the scalar field commutes with
itself at arbitrary spacelike separation.

\section{Bulk construction of local operators}

Although it won't be the main emphasis of this paper, one can
construct local bulk observables from the bulk point of view, by
solving the bulk equations of motion perturbatively
\cite{Kabat:2011rz, Heemskerk:2012mn}.  Here we sketch the
construction for scalar electrodynamics.

The equations of motion which follow from (\ref{ScalarElectrodynamics}) are
\bea
&& {1 \over \sqrt{-g}} D_M \sqrt{-g} D^M \phi = 0 \label{ScalarEOM} \\
&& {1 \over \sqrt{-g}} \partial_M \sqrt{-g} F^{MN} = J^N \label{GaugeEOM}
\eea
where $D_M = \partial_M + i q A_M$ and $J^M = iq(D^M \phi^* \phi - \phi^* D^M \phi)$.  We wish to solve these equations perturbatively
in $q$ (which we identify as being ${\cal O}(1/N)$), in the gauge $A_z = 0$.

We begin with the scalar equation of motion (\ref{ScalarEOM}), which in terms of a Christoffel connection
$\nabla$ on AdS reads
\[
\nabla_M \nabla^M \phi + i q (\nabla_M A^M) \phi + 2 i q A^M \partial_M \phi - q^2 A^2 \phi = 0
\]
We can solve this perturbatively, setting
\beas
&& \phi = \phi^{(0)} + \phi^{(1)} + \cdots \\
&& A_M = A_M^{(0)} + A_M^{(1)} + \cdots
\eeas
where
\bea
\label{phi0} \nabla_M \nabla^M \phi^{(0)} &=& 0 \\
\label{phi1} \nabla_M \nabla^M \phi^{(1)} &=& -iq(\nabla_M A^{M\,(0)}) \phi^{(0)} - 2 i q A^{M\,(0)} \partial_M \phi^{(0)} \\
\nonumber&\vdots&
\eea
The first equation can be solved -- with suitable boundary conditions
that match on to the CFT as $z \rightarrow 0$ -- using the scalar
smearing function constructed in \cite{Hamilton:2006az,
Hamilton:2006fh}.  The second equation can be solved using a
spacelike Green's function as in \cite{Kabat:2011rz,
Heemskerk:2012mn}.

Next we look at the $z$ component of the gauge field equation of
motion (\ref{GaugeEOM}), which reduces to\footnote{Indices tangent to
the boundary are raised and lowered with the Minkowski metric
$\eta_{\mu\nu}$.}
\[
\partial_z (\partial_\mu A^\mu) = - {R^2 \over z^2} J_z
\]
This fixes
\be
\label{DivA}
\partial_\mu A^\mu(x,z) = - \int_0^z dz' \, {R^2 \over z'{}^2} J_z(x,z')
\ee
In the absence of a source note that $\partial_\mu A^\mu = 0$ as in
\cite{Kabat:2012hp}, so that in fact $\nabla_M A^{M\,(0)} = 0$ in
(\ref{phi1}).

The remaining components of the gauge field equations of motion reduce to
\be
\label{phieom}
{1 \over \sqrt{-g}} \partial_M \sqrt{-g} g^{MN} \partial_N \phi_\lambda + {d - 1 \over R^2} \phi_\lambda = z \left(J_\lambda + {z^2 \over R^2} \partial_\lambda(\partial_\mu A^\mu)\right)
\ee
where we've introduced $\phi_\lambda = z A_\lambda$.  This is convenient because the left hand side of (\ref{phieom}) is the wave equation for a scalar field of mass
$m^2 R^2 = 1 - d$.  Expanding in powers of the coupling, we have the tower of equations
\bea
\label{Aeom}
&& {1 \over \sqrt{-g}} \partial_M \sqrt{-g} g^{MN} \partial_N \phi_\lambda^{(0)} + {d - 1 \over R^2} \phi_\lambda^{(0)} = 0 \\
\nonumber && {1 \over \sqrt{-g}} \partial_M \sqrt{-g} g^{MN} \partial_N \phi_\lambda^{(1)} + {d - 1 \over R^2} \phi_\lambda^{(1)} =
z J_\lambda^{(1)} - z^3 \partial_\lambda \int_0^z dz' \, {1 \over z'{}^2} J_z^{(1)}(x,z') \\
&& \nonumber \hspace{2cm} \vdots
\eea
where the first-order current $J^{(1)}$ is expressed in terms of the
lowest-order field $\phi^{(0)}$, and where we've used (\ref{DivA}) to
express $\partial_\mu A^\mu$ in terms of the bulk current.  Just as
for the scalar field, the first equation can be solved using an
appropriate scalar smearing function, while the second equation can be
solved using a spacelike Green's function.

In the rest of this paper we will see how this structure emerges
directly from the CFT, without using bulk equations of motion.

\section{CFT construction: bulk scalars}

In this section, as a warm-up illustrative example, we will see how
things work for an interacting scalar field.  This extends the
construction of \cite{Kabat:2011rz} to $d+1$ dimensions and gives results which will be useful later.  Similarly to
what was done in the AdS${}_3$ case one expects the 3-point function of
a bulk scalar and two boundary scalars to have the form
\begin{eqnarray}
<\phi_{i}(x,z){\cal O}_{j}(y_1){\cal O}_{k}(y_2)>&=&c_{ijk}\frac{1}{(y_1 -y_2)^{\Delta_{j}+\Delta_{k}-\Delta_{i}}}
\left[\frac{z}{z^2+(x-y_1)^2}\right]^{(\Delta_{j}+\Delta_{i}-\Delta_{k})/2} \nonumber \\
&\times& \left[\frac{z}{z^2+(x-y_2)^2}\right]^{(\Delta_{k}+\Delta_{i}-\Delta_{j})/2}f(\chi)
\label{3pointbulk}
\end{eqnarray}
where
\begin{equation}
\chi=\frac{[(x-y_1)^2+z^2][(x-y_2)^2+z^2]}{z^2(y_2-y_1)^2}
\end{equation}
To compute $f(\chi)$ we look at the limit of large $y_2$, where the CFT 3-point function reduces to
\begin{equation}
<{\cal O}_{i}(x){\cal O}_{j}(0){\cal O}_{k}(y_2)> \rightarrow c_{ijk}\frac{1}{(y_2)^{2\Delta_{k}}}\frac{1}{x^{2\Delta_0}}
\end{equation}
Here $x^2=\vert\vec{x}\vert^{2}-t^2$ and $\Delta_0=(\Delta_{i}+\Delta_{j}-\Delta_{k})/2$.
We now define $\phi_i(z,x)$ by smearing ${\cal O}_{i}$ into the bulk using the appropriate scalar smearing function 
\begin{equation}
\phi(z,t,\vec{x})=\frac{\Gamma(\Delta-d/2+1)}{\pi^{d/2}\Gamma(\Delta-d+1)}\int_{t'^2+\vert\vec{y}^{\,\prime}\vert^2\leq z^2}
dt'd^{d-1}y'\left(\frac{z^2-t'^2-\vert\vec{y}^{\,\prime}\vert^2}{z}\right)^{\Delta-d}{\cal O}(t+t',\vec{x}+i\vec{y}^{\,\prime})\,.
\end{equation}
This gives
\begin{equation}
<\phi_{i}(x,z){\cal O}_{j}(0) {\cal O}_{k}(y_2)>  \rightarrow c_{ijk} \frac{1}{(y_2)^{2\Delta_{k}}}g(x,z)
\end{equation}
where
\begin{equation}
g(x,z) =\frac{\Gamma(\Delta_i -\frac{d}{2}+1)}{\pi^{\frac{d}{2}} \Gamma(\Delta_i-d+1)} \int_{t'^{2}+y^{2} \leq z^2} d^{d-1}y dt'
\left(\frac{z^2-t'^{2}-y^{2}}{z}\right)^{\Delta_{i}-d}\frac{1}{((\vec{x}+i\vec{y})^2-(t+t')^{2})^{\Delta_{0}}}
\end{equation}
The integral can be done by setting all but one of the $x$ components to zero, doing the integral, and then
restoring Lorentz invariance.  The result is
\begin{equation}
g(x,z)=\frac{z^{\Delta_{i}}}{x^{2\Delta_{0}}}
F(\Delta_{0},\Delta_{0}-d/2+1,\Delta_{i}-d/2+1,-\frac{z^2}{x^2})
\label{fijk}
\end{equation}
Comparing to the large-$y_2$ behavior of (\ref{3pointbulk}) one finds 
\begin{equation}
f(\chi)=\left(\frac{\chi}{\chi-1}\right)^{\Delta_{0}}F(\Delta_{0},\Delta_{0}-\frac{d}{2}+1,\Delta_{i}-\frac{d}{2}+1,\frac{1}{1-\chi})
\label{fchi}
\end{equation}
$\chi=0$ corresponds to the bulk point being lightlike to one of the boundary points. The dangerous region is
$0<\chi <1$ where all points are spacelike to each other. We will see that the 3-point function is not analytic in this region so the operators will not commute.

The analytic structure depends on whether $d/2$ is integer or half-integer. If $d/2$ is half-integer one can use the transformation formula
\begin{eqnarray}
F(a,b,c,z)&=&(-z)^{-a}\frac{\Gamma(c)\Gamma(b-a)}{\Gamma(c-a)\Gamma(b)}F(a,a-c+1,a-b+1,\frac{1}{z})\nonumber\\
&+&(-z)^{-b}\frac{\Gamma(c)\Gamma(a-b)}{\Gamma(c-b)\Gamma(a)}F(b,b-c+1,b-a+1,\frac{1}{z})
\end{eqnarray}
This gives the scalar 3-point function a non-analytic term (near $\chi=1$) of the form
\begin{equation}
\frac{1}{(\chi-1)^{\frac{d}{2}}} \sum_{n=0} a_n (1-\chi)^{n+1}
\end{equation}
Thus we have square root singularities, which will give a non-zero commutator for $0<\chi <1$.

If $d/2$ is an integer then the above formula is not correct. Instead the correct formula to use is
\begin{eqnarray}
& &F(a,a+n,c,z)=\frac{\Gamma(c)(-z)^{-a}}{\Gamma(c-a)\Gamma(a+n)}\sum_{k=0}^{n-1}\frac{(n-k-1)!(a)_{k}(1-c+a)_{k}}{k!}(-z)^{-k}\nonumber\\
&+&\frac{\Gamma(c)(-z)^{-a}}{\Gamma(a)\Gamma(c-a-n)}\sum_{k=0}^{\infty}\frac{(a+n)_{k}(1-c+a+n)_{k}}{(n+k)! k!}[\psi(k+1)+\psi(n+k+1)\nonumber \\
&-&\psi(a+n+k)-\psi(c-a-n-k)+\ln (-z)]z^{-n-k}
\end{eqnarray}
where $\psi(x)=\frac{\Gamma^{'}(x)}{\Gamma(x)}$ and $(n)_{k}=\frac{\Gamma(n+k)}{\Gamma(n)}$.
We see that for the scalar 3-point function the terms contributing to the commutator are of the form
\begin{equation}
\sum_{k=0}^{d/2} b_{k}(1-\chi)^{-\frac{d}{2}+1+k}+\ln(\chi-1) \sum_{k=0}^{\infty} a_k (1-\chi)^{k}
\end{equation}

To summarize, for even $d$ one gets logarithmic singularities in
regions where the points are space-like separated and for odd $d$ one
gets square root singularities.

The region where there is a non-vanishing commutator between the bulk
scalar and one of the boundary scalars, while still having all three
points at bulk spacelike separation, is $0<\chi<1$.  From the above
formulas we see that the non-zero commutator in this region has the
form of a power series in $(\chi-1)$. We also see that the singularity
structure is the same regardless of the dimension of the operators
involved. We wish to define a bulk operator $\phi_{i}(x,z)$ in such a
way as to have the smallest possible commutator with the boundary
operators at spacelike separation, transform as a bulk scalar under
AdS isometries, and have the correct boundary behavior
\begin{equation}
\phi_{i}(x,z) \hspace{1mm} \stackrel{z\rightarrow 0}{\rightarrow} \hspace{1mm} z^{\Delta_{i}} {\cal O}_{i}.
\end{equation}
If we have higher dimension primary scalar operators with dimensions
$\Delta_{l}$, whose 3-point functions with ${\cal O}_{j}$ and ${\cal O}_{k}$ are non-zero, we can redefine the bulk operator
$\phi_{i}(x,z)$ to have the form
\begin{equation}
\phi_{i}(z,x)=\int dx' K_{\Delta_i}(z,x|x'){\cal O}_i(x')+\sum_{l} a_{l} \int dx' K_{\Delta_{l}}(z,x|x'){\cal O}_l(x')
\end{equation}

Since the singularity structure is the same for any ${\cal O}_{l}$, we
can choose the coefficients $a_{l}$ in such a way as to make the
commutator of order $(\chi -1)^{\Delta_{\rm max}}$ where $\Delta_{\rm
max}$ is as large as we wish. If we have an infinite number of suitable
higher dimension operators, with conformal dimensions that are
unbounded above, we can make the bulk scalar commute at bulk spacelike
separation.  This is how we define the bulk scalar field. Clearly
for any two different ${\cal O}_{j}$ and ${\cal O}_{k}$, we will need
a different tower of higher dimension primaries. Fortunately in the
large $N$ limit the required operators can be built up from operator
products of ${\cal O}_{j}$ and ${\cal O}_{k}$ with derivatives. If
${\cal O}_{j}$ and ${\cal O}_{k}$ are single trace operators this
procedure begins with a double trace operator and thus $a_{l} \sim
1/N$.\footnote{For a general discussion of large-$N$ counting in this
context see p.\ 26 of \cite{Kabat:2011rz}.}

\section{CFT construction: bulk scalars coupled to vectors}

In this section we consider charged scalar fields in the bulk and
study the corrections we need to add to the definition of a bulk
observable to take into account interactions with currents in the CFT.
There are two cases we consider.  First, in section
\ref{subsec:non-conserved}, we consider corrections due to
interactions with a non-conserved current in the CFT (dual to a
massive vector field in the bulk).  Then in section
\ref{subsec:conserved} we consider interactions with a conserved
current in the CFT (dual to a bulk gauge field).  We carry out the
construction from the CFT perspective, by adding an infinite tower of
higher-dimension operators and requiring bulk micro-causality.  Thus
we extend the program of \cite{Kabat:2011rz} to include scalars which couple
to boundary currents, conserved or not.

\subsection{Coupling to non-conserved currents\label{subsec:non-conserved}}

Following the approach of \cite{Kabat:2011rz,Kabat:2012hp} we look at the three
point function of a non-conserved current of dimension $\Delta$ and
two primary scalars of dimension $\Delta_{1}$ and $\Delta_{2}$.  Up to an overall normalization factor
the three point function is
\begin{eqnarray}
&&<j_{\mu}(x){\cal O}_{1}(y_1){\cal O}_{2}(y_2)>\nonumber\\
&=&\frac{1}{(y_1-y_2)^{\Delta_{1}+\Delta_{2}-\Delta+1}(y_1-x)^{\Delta+\Delta_1-\Delta_2-1}(y_2-x)^{\Delta+\Delta_2-\Delta_1-1}}
\left(\frac{(y_1-x)_{\mu}}{(y_1-x)^{2}}-\frac{(y_2-x)_{\mu}}{(y_2-x)^{2}}\right)\nonumber\\
\label{m3point1}
\end{eqnarray}
This can be written as
\begin{eqnarray}
&&\left(\frac{1}{\Delta_2+1-\Delta_1-\Delta}\frac{\partial}{\partial{(y_1-x)_{\mu}}}-\frac{1}{\Delta_1+1-\Delta_2-\Delta}\frac{\partial}{\partial{(y_2-x)_{\mu}}}\right)\nonumber\\
&&\left[\frac{1}{(y_1-y_2)^{\Delta_{1}+\Delta_{2}-\Delta+1}(y_1-x)^{\Delta+\Delta_1-\Delta_2-1}(y_2-x)^{\Delta+\Delta_2-\Delta_1-1}}\right]\nonumber\\
\label{m3point2}
\end{eqnarray}
where the partial derivative with respect to $(y_1-x)_{\mu}$ means we are keeping $|y_1-y_2|$ and $|y_2-x|$ fixed (and
similarly with $y_1 \leftrightarrow y_2$).
The term in square brackets in (\ref{m3point2}) can be written as
\begin{equation}
(y_2-x)^2<\tilde{\cal O}_{1}(x)\tilde{\cal O}_{2}(y_1)\tilde{\cal O}_{3} (y_2)>
\end{equation}
where the expectation value is that of three primary scalars of dimensions $\Delta,\Delta_{1},\Delta_{2}+1$ respectively.
Thus one gets
\begin{eqnarray}
&&\hspace{-1cm}<j_\mu(x)\phi_{1}(z,y_1){\cal O}_{2} (y_2)>=\left(\frac{1}{\Delta_2+1-\Delta_1-\Delta}\frac{\partial}{\partial{(y_1-x)_{\mu}}}
-\frac{1}{\Delta_1+1-\Delta_2-\Delta}\frac{\partial}{\partial{(y_2-x)_{\mu}}}\right)\nonumber\\
&&\hspace{-1cm}\frac{1}{(x -y_2)^{\Delta+\Delta_{2}-1-\Delta_{1}}}\left[\frac{z}{z^2+(y_1-x)^2}\right]^{(\Delta+\Delta_{1}-\Delta_{2}-1)/2}
\left[\frac{z}{z^2+(y_1-y_2)^2}\right]^{(\Delta_{2}+1+\Delta_{1}-\Delta)/2}\nonumber \\
&&\hspace{-1cm}\left(\frac{\chi}{\chi-1}\right)^{\Delta_{0}}F(\Delta_{0},\Delta_{0}-\frac{d}{2}+1,\Delta_{1}-\frac{d}{2}+1,\frac{1}{1-\chi})
\end{eqnarray}
where
\begin{equation}
\chi=\frac{[(x-y_1)^2+z^2][(y_2-y_1)^2+z^2]}{z^2(y_2-x)^2}
\end{equation}
and $\Delta_{0}=\frac{1}{2}(\Delta_{1}+\Delta-\Delta_{2}-1)$.

We know the singularity structure of the scalar three point function
(\ref{3pointbulk}), and we know we can cancel the non-causal
singularities in it by adding higher dimension smeared scalar
primaries. Thus when smearing ${\cal O}_{1}$ into the bulk we can cancel
the non-causal singularities in (\ref{m3point1}) by adding a tower of
higher dimension smeared scalar primaries to our definition of a bulk
scalar.  This should come as no surprise since from the bulk point of
view a theory of a massive vector coupled to scalars is a conventional
local theory.  Thus there should be no obstacle to constructing a
local bulk scalar field.

However the question remains, where do these higher dimension scalars
come from?  In the large $N$ limit we can construct them as double
trace operators.  For example the first higher dimension operator we
can construct (starting from the case $\Delta_{1}=\Delta_{2}$) is
\begin{equation}
\alpha \partial_{\mu} j^{\mu} {\cal O}_{2} + \beta  j^{\mu}\partial_{\mu}{\cal O}_{2}
\label{higherscalar}
\end{equation}
With the choice $\alpha=\frac{1}{\Delta-d+1}$ and
$\beta=-\frac{1}{\Delta_{2}}$ this is a primary scalar.  This
reproduces the bulk result, where for a massive vector field the first
correction is sourced by terms proportional to
$A^{M} \partial_{M}\phi$, since the near boundary behavior of this
bulk quantity is exactly the operator above.\footnote{The $z$
component of $A^{M} \partial_{M}\phi$ gives rise to the
$\partial_{\mu} j^{\mu} {\cal O}_{2}$ term, while the $\mu$
components give rise to $j^{\mu}\partial_{\mu}{\cal O}_{2}$.  This
follows from the massive vector smearing function given below in
(\ref{MassiveVectorSmearingFunction}), (\ref{MassiveVectorSmearingFunction1}).}  Additional higher dimension
operators can be constructed by inserting derivatives in various
fashions.

\subsection{Coupling to conserved currents\label{subsec:conserved}}

Let's see how things change when we have a conserved current in the CFT, dual to a gauge field in the bulk.  From the CFT point of view the only
difference is that now there is a Ward identity which restricts correlation functions involving the current, e.g.
\be
\label{Ward}
\partial_\mu \langle j^\mu(x) {\cal O}(y_1) \bar{\cal O}(y_2) \rangle = -iq \langle {\cal O}(y_1) \bar{\cal O}(y_2) \rangle \left(\delta(x - y_1) - \delta(x - y_2)\right)
\ee
Here $q$ is the charge of the scalar operator and an overbar denotes complex conjugation.

We start with the three point function of a conserved current (which necessarily has dimension $d - 1$)
and two primary scalars having the same dimension $\Delta_1$.
\begin{equation}
<j_{\mu}(x){\cal O}(y_1)\bar{\cal O}(y_2)>=\frac{1}{(y_1-y_2)^{2\Delta_{1}-d+2}(y_2-x)^{d-2}(y_1-x)^{d-2}}\left(\frac{(y_1-x)_{\mu}}{(y_1-x)^{2}}-\frac{(y_2-x)_{\mu}}{(y_2-x)^{2}}\right)
\label{jphiphi}
\end{equation}
As we saw before in (\ref{m3point2}) this can be written as
\begin{eqnarray}
\frac{1}{2-d}\left(\frac{\partial}{\partial_{(y_1-x)_{\mu}}}-\frac{\partial}{\partial_{(y_2-x)_{\mu}}}\right)
\left[\frac{1}{(y_1-y_2)^{2\Delta_{1}-d+2}(y_1-x)^{d-2}(y_2-x)^{d-2}}\right]\nonumber\\
\label{3pointcurrent}
\end{eqnarray}
and the term in the square bracket is just 
\begin{equation}
(y_2-x)^2<\tilde{\cal O}_1(x)\tilde{\cal O}_2(y_1)\tilde{\cal O}_3(y_2)>
\end{equation}
where the three primary scalars are of dimension 
\begin{equation}
\tilde\Delta_{1}=d-1 \hspace{8mm} \tilde\Delta_{2}=\Delta_{1} \hspace{8mm} \tilde\Delta_{3}=\Delta_{1}+1
\end{equation}
One can use the result (\ref{3pointbulk}) to find 
\begin{eqnarray}
& &<j_{\mu}(x)\phi(z,y_1)\bar{\cal O}(y_2)>=\frac{1}{2-d}\left(\frac{\partial}{\partial_{(y_1-x)_{\mu}}}-\frac{\partial}{\partial_{(y_2-x)_{\mu}}}\right)\times \nonumber\\
& & \frac{1}{(y_2 - x)^{d-2}}\left[\frac{z}{z^2+(x-y_1)^2}\right]^{(d-2)/2} 
\left[\frac{z}{z^2+(y_1-y_2)^2}\right]^{(\Delta_1-d/2+1)}\left(\frac{\chi}{\chi-1}\right)^{(d-2)/2}\nonumber\\
\label{gaugess}
\end{eqnarray}
where now 
\begin{equation}
\chi=\frac{[(x-y_1)^2+z^2][(y_1-y_2)^2+z^2]}{z^2(y_2-x)^2}
\end{equation}
However at this point we cannot proceed as we did for a non-conserved
current: the Ward identity (\ref{Ward}) forbids a non-zero three point
function of a conserved current with two scalars of unequal dimension,
which means we cannot correct our definition of a bulk scalar by
adding higher-dimension primaries.  This can also be seen from the
results of the previous section, where the higher-dimension primary we
had to add in the non-conserved case (\ref{higherscalar}) involved the
divergence of the current.\footnote{One could try to take a limit
where the current is conserved, with $\Delta \rightarrow d - 1$, but
then in (\ref{higherscalar}) we'd have $\alpha \rightarrow
\infty$ while the divergence of the current goes to zero. It does
not seem to make sense to take such a limit at the operator level.}

More specifically in the limit of large $y_2$ the leading term in (\ref{jphiphi}) is
\begin{equation}
\frac{1}{y_{2}^{2\Delta_1}}\frac{(y_1-x)_{\mu}}{(y_1-x)^{d}}
\label{type1}
\end{equation}
Upon smearing ${\cal O}(y_1)$ into the bulk this generates some non-causal terms we would like to cancel. We could try to
correct our definition of the bulk scalar by adding a smeared $j_{\rho}\partial^{\rho}{\cal O}(y_1)$, but this won't
work since by large-$N$ factorization
\[
\langle j_\mu(x) \, j_\rho \partial^\rho {\cal O}(y_1) \, \bar{\cal O}(y_2) \rangle = \langle j_\mu(x) \, j_\rho(y_1)\rangle
\, \langle \partial^\rho{\cal O}(y_1) \bar{\cal O}(y_2)\rangle
\]
and thus as $y_2 \rightarrow \infty$ the leading dependence on $y_2$ that such a correction would produce is
\begin{equation}
\frac{y_{2\rho}}{y_{2}^{2\Delta_1+2}}
\label{type2}
\end{equation}
This means there is no way to cancel the unwanted terms. A term like
(\ref{type1}) could be canceled in the non-conserved case, by adding a
correction of the form $\partial^{\rho}j_{\rho}{\cal O}(y_1)$; with no
derivatives acting on ${\cal O}$ the leading $y_2 \rightarrow \infty$
dependence would be $1/y_2^{2 \Delta_1}$.  But for conserved currents
this operator is not available.  From this perspective this is the
only difference (though a crucial one) between conserved and
non-conserved currents.

This failure to restore bulk locality is actually desirable, since as
we will show in section \ref{subsec:Gauss} the resulting
non-commutativity is exactly what one needs in order to satisfy the
bulk Gauss constraint.  But it raises a question, how should we go
about determining the appropriate higher-dimension operators to add to
our bulk scalar?  To find the answer we make a strategic retreat, and
study causality for correlators involving the field strength of a massive vector.

\goodbreak\noindent
{\em Causality for massive vector field strengths} \\
Consider the 3-point function of two scalars and one field
strength $F_{\mu \nu}=\partial_{\mu}j_{\nu}-\partial_{\nu}j_{\mu}$ built from a non-conserved current.
From (\ref{m3point1}) this is
\begin{eqnarray}
& &<F_{\mu \nu}(x){\cal O}_{1}(y_1){\cal O}_{2}(y_2)> \sim \nonumber\\
& &  \frac{1}{(y_1-y_2)^{\Delta_{1}+\Delta_{2}-\Delta+1}(y_1-x)^{\Delta+\Delta_1-\Delta_2+1}(y_2-x)^{\Delta+\Delta_2-\Delta_1+1}}\left[(y_1-x)_{\mu}(y_2-x)_{\nu}- \mu \leftrightarrow \nu \right]
\nonumber \\
\label{f3point}
\end{eqnarray}
As shown in section \ref{subsec:non-conserved}, when lifted into the bulk
this correlator has non-causal singularities which can be canceled by adding
higher-dimension operators.

For example, when the higher dimension operator (\ref{higherscalar})
is inserted in place of ${\cal O}_1$ in the original 3-point function,
the resulting CFT correlator can be computed by large-$N$
factorization as a product of two-point functions.  From the current -
current correlator\footnote{Some useful formulas are recorded in
section 4.1 of \cite{Kabat:2012hp}.}
\begin{equation}
<j_{\mu}(x) j_{\nu}(0)>=\left(\eta_{\mu \nu}-\frac{2x_{\mu}x_{\nu}}{x^2}\right)\frac{1}{(x^2)^{\Delta}}
\end{equation}
it follows that
\begin{equation}
<F_{\mu \nu}(x) \partial^{\rho}j_{\rho}>=0
\end{equation}
and therefore at leading order in $1/N$
\begin{equation}
< F_{\mu \nu}(x) \, \partial^\rho j_\rho {\cal O}_2(y_1) \, {\cal O}_2(y_2) > \, \simeq \, 0
\end{equation}
To leading order in $1/N$ the terms that are missing for conserved currents do not contribute in the non-conserved case, at least for a 3-point
function involving $F_{\mu\nu}$.  Since we know the massive vector can be made local in the bulk, this means the cancellation must come from
the $j_\rho \partial^\rho {\cal O}_2$ term.  To leading order in $1/N$ this term gives
\begin{equation}
<F_{\mu \nu}(x) j_{\rho}(y_1)><\partial^{\rho}{\cal O}_2(y_1){\cal O}_2(y_2)>
\label{gcorrection1}
\end{equation}
and indeed after some algebra this has the same form as
(\ref{f3point}), with $\Delta_{1}=\Delta+\Delta_{2}+1$ as appropriate
for this operator. This result is valid even in the limit $\Delta
\rightarrow d - 1$ where the current is conserved, since no property
of the non-conserved current was used (i.e.\ having a divergence of
the current $\partial^{\mu}j_{\mu} \neq 0$ played no role).

\noindent
{\em Lessons for massless vectors} \\
For massive vectors we found that the operator which is absent in the
conserved case, namely $\partial^\rho j_\rho {\cal O}$, played no role
in restoring causality for correlators involving the boundary field strength $F_{\mu\nu}$.
Thus we expect that in a three point function of the type $\langle F_{\mu\nu} {\cal O} \bar{\cal O} \rangle$
locality can be respected even for conserved currents, just by adding
smeared operators which are scalars but not primary.  For example,
given the operator (\ref{higherscalar}), we would correct the bulk
scalar by just smearing the $j_{\rho}\partial^{\rho}{\cal O}$ term.
As shown below (\ref{gcorrection1}) this suffices to restore causality
for correlators involving a massless boundary field strength.

So for conserved currents, even though one cannot build a primary
scalar out of the available ingredients (the current, other primary
scalars, derivative operators), one can still build an operator which
can be treated as though it were a primary scalar, at least when
inserted in three point functions involving $F_{\mu \nu}$ rather than
$j_{\mu}$. By taking this non-primary operator and smearing it as
though it were primary we can cancel the non-local terms in $\langle
F_{\mu\nu} {\cal O} \bar{\cal O} \rangle$, just as we did for
non-conserved currents.  In this way the bulk scalar can be corrected
so that it is local with respect to the field strength $F_{\mu\nu}$ on
the boundary.\footnote{We do not expect it to be local with respect to $F_{\mu z}
\sim j_\mu$ near the boundary.}    This fits with the bulk perspective developed in section \ref{sect:bulk},
where at equal times the bulk scalar commuted with the field strengths $\pi_{\hat{\imath}}$, $F_{\hat{\imath}\hat{\jmath}}$.  This requirement is what captures
the appropriate notion of micro-causality when there are conserved
currents.

Presumably all of the required higher-dimension non-primary scalar
operators can be constructed in the $1/N$ expansion.  For example,
let's look at the next correction to (\ref{higherscalar}), involving
an operator of dimension $d + \Delta_2 + 2$.  To determine its form we
first build a scalar primary out of a non-conserved current of
dimension $\Delta$, then we drop terms involving the divergence of the
current and take the limit $\Delta \rightarrow d-1$ with $\Delta_1 =
\Delta_2$.  Denoting the scalar operators ${\cal O}$ and $\bar{\cal O}$,
this leads to
\begin{equation}
\alpha (\nabla^2 j_{\rho})\partial^{\rho}{\cal O} + \beta j_{\rho}\partial^{\rho}\nabla^2 {\cal O} +\gamma \partial_{\delta} j_{\rho}\partial^{\rho}\partial^{\delta} {\cal O} 
\end{equation}
with
\be
\alpha=\frac{1}{2d^2} \hspace{8mm}
\beta=\frac{1}{2(\Delta_1+1)(2\Delta_1+2-d)} \hspace{8mm}
\gamma=-\frac{1}{2d(\Delta_1+1)}
\ee

Adding these higher-dimension non-primary operators cancels the
unwanted non-analyticity and restores locality in correlators
involving $F_{\mu \nu}$.  This means the corrected bulk scalar
will commute at spacelike separation with $F_{\mu\nu}$ on the
boundary.  It also means that two scalar fields will commute at
spacelike separation, even in the presence of a spectator $F_{\mu \nu}$.

\subsection{AdS covariance}

The procedure we have outlined restores bulk locality, at least in
correlators involving $F_{\mu\nu}$, but it seems dangerous.  We have
added to the original scalar field an operator which is smeared like a
primary scalar but is not actually a primary scalar.  This means the
resulting bulk field will not transform as a bulk scalar under AdS
isometries.  At first this sounds problematic, but we now show that it's
the expected result: in holographic gauge charged scalar fields
acquire an anomalous transformation rule under AdS isometries which do
not preserve the gauge-fixing condition.

First let's study this from the bulk point of view.  We are working in
holographic gauge, $A_{z}=0$. This completely fixes the gauge, so all
our bulk fields are physical.  In this gauge a charged scalar
$\phi(z,x)$ can be identified with the manifestly gauge-invariant
observable
\begin{equation}
\phi_{\rm phys}(z,x)=e^{i\int dz A_{z}} \phi(z,x)
\end{equation}
As such, under an isometry which does not preserve the condition $A_z
= 0$ the field will not transform like a scalar: rather a compensating
gauge transformation will be required.  Indeed under a special
conformal transformation
\begin{equation}
\phi'_{\rm phys}(z',x')=e^{i\lambda (z,x)}\phi_{\rm phys}(z,x)
\end{equation}
where $\lambda (z,x)$ is given to first order in the parameter of the
special conformal transformation $b_{\mu}$ by \cite{Kabat:2012hp}
\begin{equation}
\lambda=-\frac{1}{{\rm vol}(S^{d-1})}\int d^{d}x' \theta(\sigma z')2b\cdot j
\end{equation}
We will be interested in the boundary behavior
\begin{equation}
\lambda(z,x) \hspace{1mm} \stackrel{z\rightarrow 0}{\rightarrow} \hspace{1mm} z^d 2b \cdot j
\end{equation}
from which we find that to first order in $b_{\mu}$ and as $z\rightarrow 0$
\begin{equation}
\phi'_{\rm phys}(z',x')\stackrel{z\rightarrow 0}{\rightarrow}\phi_{\rm phys}(z,x)+2iz^{\Delta+d}b\cdot j {\cal O}(x)
\label{pstrans}
\end{equation}

Now let's see how the CFT reproduces this behavior.  We saw that to
leading order in $1/N$ and as $z \rightarrow 0$ the correction to the
bulk scalar field in the CFT has the form
\begin{equation}
\phi(z,x)=\int K_{\Delta}(z,x|y){\cal O}(y)+\int K_{\Delta+d}(z,x|y)j^{\mu}\partial_{\mu}{\cal O}(y)
\label{pstrans2}
\end{equation}
We use the behavior under infinitesimal special conformal transformations acting on $y$
\begin{eqnarray}
d^d y' & = & (1+2d b\cdot y)d^d y \nonumber \\
K'_{\Delta} & = & K_{\Delta}(1+2(\Delta-d) b\cdot y) \nonumber \\
{\cal O}'(y') & = & (1-2\Delta b\cdot y) {\cal O}(y) \nonumber \\
j'{}^{\mu}(y') & = & (1-2d b\cdot y)\frac{\partial y'{}^{\mu}}{\partial y^{\nu}}j^{\nu}(y)
\end{eqnarray}
The right hand side of (\ref{pstrans2}) transforms to
\be
\int K_{\Delta}(z,x|y){\cal O}(y)+\int K_{\Delta+d}(z,x|y)j^{\mu}\partial_{\mu}{\cal O}(y)-\Delta\int K_{\Delta+d}(z,x|y)2b\cdot j{\cal O}(y)
\ee
The last term as $z \rightarrow 0$ behaves like
\begin{equation}
z^{\Delta+d}b\cdot j {\cal O}(x)
\end{equation}
which matches what one expects from the bulk perspective. So the
inability to construct a higher-dimension primary scalar from a
conserved current in the CFT, translates in a nice way to the
anomalous behavior under AdS isometries of a charged scalar field in
the bulk.

\subsection{Gauss constraint\label{subsec:Gauss}}

Although we have been able to correct our definition of a bulk scalar
field so as to achieve locality in correlators involving the boundary
field strength $F_{\mu\nu}$, correlators involving the conserved
current $j_\mu$ itself will still be non-local.  We now show that this
was to be expected, since the non-locality which is present is exactly
the bulk non-locality required by the Gauss constraint.

We start with the 3-point function (\ref{gaugess}) of a bulk scalar, a
boundary conserved current and an additional boundary scalar
\begin{eqnarray}
& & <\phi(z,y_1) j_{\mu}(x) \bar{\cal O}(y_2)> = \frac{1}{2-d}\left(\frac{d}{d(y_1-x)_{\mu}}-\frac{d}{d(y_2-x)_{\mu}}\right)\times\label{gss}\\
& & \nonumber \left[ \frac{z^{\Delta_{1}}}{(y_2-x)^{d-2}}\left( \frac{1}{z^2+(y_1-y_2)^2}\right)^{\Delta_1-\frac{d-2}{2}}
\left(z^2+(y_1-x)^2-\frac{z^2(y_2-x)^2}{z^2+(y_1-y_2)^2}\right)^{\frac{2-d}{2}}\right]
\end{eqnarray}
We assume the points $x$ and $y_2$ are spacelike to each other. We
compute the commutator of the current and the bulk operator inside the
3-point function as the difference of two $i \epsilon$ prescriptions,
one where the time component of $x$ has a $+i\epsilon$ and one where
it has a $-i\epsilon$. The only singularities that can contribute to
the commutator arise when the derivatives act on the third factor in
(\ref{gss}).  These derivatives give
\begin{equation}
2\left(z^2+(y_1-x)^2-\frac{z^2(y_2-x)^2}{z^2+(y_1-y_2)^2}\right)^{\frac{-d}{2}}((y_1-x)_{\mu}-\frac{(y_2-x)_{\mu}z^2}{z^2+(y_1-y_2)^2})
\label{gauss2}
\end{equation}
Note that
\begin{equation}
z^2+(y_1-x)^2-\frac{z^2(y_2-x)^2}{z^2+(y_1-y_2)^2}=\frac{(y_1 -y_2)^2}{z^2+(y_1 -y_2)^2}\left(x-y_1-\frac{z^2(y_1-y_2)}{(y_1-y_2)^2}\right)^2
\end{equation}
and that the delta function in $d-1$ dimensions can be written as
\begin{equation}
\delta(\vec{x})=\lim_{\epsilon \rightarrow 0}\frac{\Gamma(d/2)}{\pi^{d/2}}\frac{\epsilon}{(\vert\vec{x}\vert^2+\epsilon^2)^{d/2}}.
\end{equation}
In the simple case (one can generalize this) that the time components of $x$, $y_1$, $y_2$ are equal, then taking the
difference of (\ref{gauss2}) with $+i\epsilon$ and $-i\epsilon$ prescriptions gives zero for $\mu \neq 0$, while for
$\mu=0$ we get
\begin{equation}
\frac{2i\pi^{d/2}}{\Gamma(d/2)}\left[\frac{(y_1 -y_2)^2}{z^2+(y_1 -y_2)^2}\right]^{-\frac{d}{2}+1}\delta(x-y_1-\frac{z^2(y_1-y_2)}{(y_1-y_2)^2})
\end{equation}
To find the commutator with the charge operator $Q$ we restore the first two factors in (\ref{gss}) and
integrate over the spatial coordinates $\vec{x}$. This gives
\begin{equation}
<[\phi(y_1,z) , Q] \bar{\cal O}(y_2)> \,\,\sim \left[\frac{z}{z^2+(y_1-y_2)^2}\right]^{\Delta_1}\sim\,\, <\phi(z,y_1) \bar{\cal O}(y_2)>
\label{commutator}
\end{equation}
which is the expected commutator of the charge operator with a charged scalar field.  This shows that the bulk Gauss constraint is obeyed,
at least when the lowest-order smearing function is used.  It would be interesting to show that (\ref{commutator}) continues to hold when higher-dimension
operators are added to the definition of the bulk scalar field.

\subsection{Scalar commutator}

Adding a higher dimension non-primary operator to our definition of a
bulk scalar field had some desirable properties: it made correlators
with $F_{\mu\nu}$ local, and it gave the scalar field the correct
behavior under AdS isometries.  However one might wonder: does the
resulting scalar field commute with its complex conjugate at bulk
spacelike separation?  From the bulk perspective developed in section
\ref{sect:bulk} we would expect this to happen even in the presence of
interactions.  Here we give some evidence for this from the CFT point
of view.

As shown in section \ref{subsec:conserved}, our bulk scalars still
commute inside a 3-point function with a boundary field strength
$F_{\mu \nu}$, so let's examine what happens in a 3-point function
with a gauge field.  This was given in (\ref{jphiphi}). The leading
$y_2 \rightarrow \infty$ singularity of this expression cannot be
canceled, but in order to isolate the commutator between the bulk
scalar and the boundary scalar, we instead look at the limit $x
\rightarrow \infty$.  In this limit
\begin{equation}
<j_{\mu}(x){\cal O}(y_1) \bar{\cal O}(y_2)> \hspace{1mm} \stackrel{x \rightarrow \infty}{\sim} \hspace{1mm}
\frac{1}{x^{2d-2}}(\eta_{\mu \nu}-\frac{2x_{\mu}x_{\nu}}{x^2}) \partial^{\nu}_{y_1}\frac{1}{(y_1-y_2)^{2\Delta_{1}-d}}
\end{equation}
Smearing the first scalar operator into the bulk we find
\begin{equation}
<j_{\mu}(x) \phi(z,y_1) \bar{\cal O}(y_2)> \hspace{1mm} \stackrel{x \rightarrow \infty}{\sim} \hspace{1mm}
\frac{1}{x^{2d-2}}(\eta_{\mu \nu}-\frac{2x_{\mu}x_{\nu}}{x^2})\frac{z^{\Delta_1}(y_1-y_2)^{\nu}}{(y_1-y_2)^{d}((y_1-y_2)^2+z^2)^{\Delta_1-d+1}}
\end{equation}
The leading singularity of this expression as $(y_1 - y_2)^2 \rightarrow 0$ is
\begin{equation}
\frac{1}{x^{2d-2}}(\eta_{\mu \nu}-\frac{2x_{\mu}x_{\nu}}{x^2})\frac{(y_1-y_2)^{\nu} z^{2d-2-\Delta_1}}{(y_1-y_2)^{d}}
\label{leadsing}
\end{equation}

Now let's examine the leading behavior as $x \rightarrow \infty$ when
we insert a non-primary operator of the type we talked about. We
consider operators of the form\footnote{These are not the only
corrections, but these are the operators which contribute to the
leading behavior as $x \rightarrow \infty$.}
\begin{equation}
j_{\nu}\partial^{\nu}(\nabla^{2})^{n}{\cal O}(y_1)
\end{equation}
Inserting this operator instead of ${\cal O}(y_1)$ in the 3-point
function gives the leading $x \rightarrow \infty$ behavior of the CFT
correlator
\begin{equation}
\frac{1}{x^{2d-2}}(\eta_{\mu \nu}-\frac{2x_{\mu}x_{\nu}}{x^2})\partial^{\nu}_{y_1}\frac{1}{(y_1-y_2)^{2\Delta_1+2n}}
\end{equation}
Smearing $y_1$ into the bulk with a scalar smearing function of
dimension $\Delta_1+d+2n$ gives the large-$x$ behavior
\begin{equation}
\frac{1}{x^{2d-2}}(\eta_{\mu \nu}-\frac{2x_{\mu}x_{\nu}}{x^2})\partial^{\nu}_{y_1}\frac{z^{\Delta_1+d+2n}}{(y_1-y_2)^{2\Delta_1+2n}}
F(\Delta_1+n,\Delta_1+n-\frac{d}{2}+1,\Delta_1+\frac{d}{2}+2n+1,-\frac{z^2}{(y_1-y_2)^{2}})
\end{equation}
Using the identity
\begin{equation}
x\frac{d}{dx}F(a,b,c,x)=a(F(a+1,b,c,x)-F(a,b,c,x))
\end{equation}
this can be rewritten as 
\begin{equation}
\frac{1}{x^{2d-2}}(\eta_{\mu \nu}-\frac{2x_{\mu}x_{\nu}}{x^2})\frac{(y_1-y_2)_{\mu}z^{\Delta_1+d+2n}}{(y_1-y_2)^{2\Delta_1+2n+2}}
F(\Delta_1+n+1,\Delta_1+n-\frac{d}{2}+1,\Delta_1+\frac{d}{2}+2n+1,-\frac{z^2}{(y_1-y_2)^{2}})
\end{equation}

The leading singularity of this expression as $(y_1-y_2)^2 \rightarrow
0$ matches (\ref{leadsing}).  With enough such higher dimension
operators one can cancel the non-analyticity to any order in
$\frac{(y_1-y_2)^2}{z^2}$. (Note that in this limit the problematic
regime is $-1<\frac{(y_1-y_2)^2}{z^2}<0$.)  While this is not a
complete proof that adding higher dimension non-primary operators
makes the bulk scalar field commute with itself inside a 3-point
function with a gauge field, it is a strong indication of it.

\section{CFT construction: bulk vectors}

In this section we look at the correction, from the CFT perspective,
that one needs to add to lift a boundary current to an interacting
local vector field in the bulk.  We first consider non-conserved
currents in the CFT, dual to massive vectors in the bulk, then treat
conserved currents.

\subsection{Bulk massive vectors}

We begin by computing the three-point function of a massive vector in
the bulk with two scalars on the boundary.  Then we look at the
higher-dimension operators we need to add to cancel the unwanted
singularities in this expression.

Our starting point is the three-point function of a boundary current
of dimension $\Delta$ with two scalar operators of dimension
$\Delta_1$ in a $d$-dimensional CFT.
\begin{eqnarray}
&&<j_{\mu}(x){\cal O}(y_1)\bar{\cal O}(y_2)> \nonumber\\
&&=\frac{1}{(y_1-y_2)^{2\Delta_{1}-\Delta+1}(y_1-x)^{\Delta-1}(y_2-x)^{\Delta-1}}
\left(\frac{(y_1-x)_{\mu}}{(y_1-x)^{2}}-\frac{(y_2-x)_{\mu}}{(y_2-x)^{2}}\right)\nonumber\\
\label{m3point1a}
\end{eqnarray}
For  $A_{z}^{0}(x)=\frac{1}{d-\Delta-1}\partial^{\mu}j_{\mu}(x)$ the correlator is
\begin{eqnarray}
<A_{z}^{0}(x){\cal O}(y_1)\bar{\cal O}(y_2)>&=& -\frac{1}{(y_1-y_2)^{2\Delta_{1}-\Delta+1}(y_1-x)^{\Delta+1}(y_2-x)^{\Delta-1}}\nonumber\\
&+&\frac{1}{(y_1-y_2)^{2\Delta_{1}-\Delta+1}(y_1-x)^{\Delta-1}(y_2-x)^{\Delta+1}}\nonumber
\end{eqnarray}
The two terms on the right are the three-point functions of scalar
operators of dimensions $(\Delta,\Delta_{1}+1,\Delta_{1})$ and
$(\Delta,\Delta_{1},\Delta_{1}+1)$ respectively.\footnote{This is a
simplification that only occurs when the two scalar operators are of
the same dimension, but since this is the interesting case for a
conserved current we only treat this case.}

In \cite{Kabat:2012hp} the smearing function for uplifting a non-conserved primary current of dimension $\Delta$
to a massive vector field in the bulk was found to be
\bea
zA_{\mu}(z,x) = \int K_{\Delta}(z,x,x')j_{\mu}(x') +\frac{z}{2(\Delta-d/2+1)}\int K_{\Delta+1}(z,x,x')\partial_{\mu}A_{z}^{0}(x') \nonumber\\
\label{MassiveVectorSmearingFunction}
\eea
for the $\mu$ components and
\begin{equation}
A_{z}(z,x)= \int K_\Delta(z,x,x') A_{z}^{0}(x')
\label{MassiveVectorSmearingFunction1}
\end{equation}
for the $z$ component.  Here
$A_{z}^{0}(x)=\frac{1}{d-\Delta-1}\partial^{\mu}j_{\mu}(x)$ and
$K_{\Delta}(z,x,x')$ is the scalar smearing function appropriate for a
scalar primary of dimension $\Delta$.  Since $A_{z}$ is smeared into
the bulk with a scalar smearing function, we can borrow our scalar
result (\ref{3pointbulk}), (\ref{fchi}) to get
\begin{eqnarray}
&&\hspace{-1cm}<A_{z}(x,z){\cal O}(y_1)\bar{\cal O}(y_2)>\nonumber\\
&&\hspace{-1cm}=\frac{-1}{(y_1-y_2)^{2\Delta_1 +1-\Delta} }\left(\frac {\chi}{\chi-1}\right)^{\frac{\Delta+1}{2}}F(\frac{\Delta+1}{2},\frac{\Delta-d+3}{2},\Delta-\frac{d}{2}+1,\frac{1}{1-\chi}) \nonumber\\
&&\hspace{-1cm}\left[\left(\frac{z}{z^2+(x-y_1)^2}\right)^{\frac{\Delta+1}{2}}\left(\frac{z}{z^2+(x-y_2)^2}\right)^{\frac{\Delta-1}{2}}-\left(\frac{z}{z^2+(x-y_1)^2}\right)^{\frac{\Delta-1}{2}}
\left(\frac{z}{z^2+(x-y_2)^2}\right)^{\frac{\Delta+1}{2}}\right]\nonumber
\end{eqnarray}
where
\begin{equation}
\chi = \frac{[(x-y_1)^2+z^2][(x-y_2)^2+z^2]}{z^2(y_1-y_2)^2}
\end{equation}
is invariant under conformal transformations.  After some algebra and using
\begin{equation}
F(a,b,c,x)=(1-x)^{c-a-b}F(c-a,c-b,c,x)
\end{equation}
this can be written as the $z$ component of the quantity
\begin{eqnarray}
&&\hspace{-1cm}<A_{M}(x,z){\cal O}(y_1) \bar{\cal O}(y_2)> \label{mvss} \\
&&\hspace{-1cm} \equiv \frac{-1}{2(y_1-y_2)^{2\Delta_1}(\chi-1)^{\frac{\Delta-1}{2}}}F(\frac{\Delta-1}{2},\frac{\Delta-d+1}{2},\Delta-\frac{d}{2}+1,\frac{1}{1-\chi})
\partial_{M}^{x}\ln \frac{(x-y_1)^2 +z^2}{(x-y_2)^2 +z^2}\nonumber
\end{eqnarray}
where $M$ is a vector index in the bulk.

Although we've only calculated the $z$ component, this must be the
complete result, since (\ref{mvss}) has the correct behavior under
conformal transformations to represent the three-point function of a
bulk massive vector with two boundary scalars.  But as a check of this
result, and to develop some formulas that will be useful in the
sequel, we now show that (\ref{mvss}) gives the correct $y_2
\rightarrow \infty$ asymptotic behavior for the $\mu$ components of
the bulk vector.

The leading behavior of (\ref{m3point1a}) as $y_2 \rightarrow \infty$ is
\begin{equation}
<j_{\mu}(x){\cal O}(0)\bar{\cal O}(y_2)> \,\,\sim \frac{1}{y_{2}^{2\Delta_1}}\frac{1}{1-\Delta}\partial_{\mu}\frac{1}{x^{\Delta-1}}
\label{jOOlargey2}
\end{equation}
We will also need the leading behavior
\begin{equation}
<A^{0}_{z}(x){\cal O}(0)\bar{\cal O}(y_2)> \,\,\sim \frac{1}{y_{2}^{2\Delta_1}}\frac{1}{x^{\Delta+1}}
\end{equation}
Using the massive vector smearing function (\ref{MassiveVectorSmearingFunction}) one finds in the large $y_2$ limit
\begin{equation}
<zA_{\mu}(z,x){\cal O}(0)\bar{\cal O}(y_2)>\,\,\sim \frac{1}{y_{2}^{2\Delta_1}}\frac{\Gamma(\Delta-d/2+1)}{\pi^{d/2}\Gamma(\Delta-d+1)}\partial_{\mu}
\left(\frac{I_{1}}{1-\Delta}+\frac{z I_{2}}{2(\Delta-d+1)}\right)
\label{largey2AOO}
\end{equation}
where
\begin{eqnarray}
I_{1} &=& \int_{t'^{2}+y^{2} \leq z^2} d^{d-1}y dt'\left(\frac{z^2-t'^{2}-y^{2}}{z}\right)^{\Delta-d}\frac{1}{((\vec{x}+i\vec{y})^2-(t+t')^{2})^{\frac{\Delta-1}{2}}} \nonumber \\
I_{2} &=& \int_{t'^{2}+y^{2} \leq z^2} d^{d-1}y dt'\left(\frac{z^2-t'^{2}-y^{2}}{z}\right)^{\Delta-d+1}\frac{1}{((\vec{x}+i\vec{y})^2-(t+t')^{2})^{\frac{\Delta+1}{2}}} \nonumber
\end{eqnarray}
These integrals give
\begin{eqnarray}
I_{1} &=& \frac{{\rm vol}(S^{d-2})\Gamma(\frac{d-1}{2})\Gamma(\frac{1}{2})\Gamma(\Delta-d+1)}{2\Gamma(\Delta-\frac{d}{2}+1)}
\frac{z^{\Delta}}{x^{\Delta-1}}F(\frac{\Delta-1}{2},\frac{\Delta-d+1}{2},\Delta-\frac{d}{2}+1,-\frac{z^2}{x^2})\nonumber \\
I_{2} &=& \frac{{\rm vol}(S^{d-2})\Gamma(\frac{d-1}{2})\Gamma(\frac{1}{2})\Gamma(\Delta-d+2)}{2\Gamma(\Delta-\frac{d}{2}+2)}
\frac{z^{\Delta+1}}{x^{\Delta+1}}F(\frac{\Delta+1}{2},\frac{\Delta-d+3}{2},\Delta-\frac{d}{2}+2,-\frac{z^2}{x^2})\nonumber \\
\label{i1i2}
\end{eqnarray}
Putting this all together we find
\begin{eqnarray}
<zA_{\mu}(z,x){\cal O}(0)\bar{\cal O}(y_2)> \sim \frac{1}{y_{2}^{2\Delta_1}}\frac{1}{1-\Delta}\partial_{\mu}\left[\frac{z^{\Delta}}{x^{\Delta-1}}
F(\frac{\Delta-1}{2},\frac{\Delta-d+3}{2},\Delta-\frac{d}{2}+1,-\frac{z^2}{x^2})\right]\nonumber \\
\label{massivevector3point}
\end{eqnarray}
Using the hypergeometric identity
\begin{equation}
x\frac{d}{dx}F(a,b,c,x)=a(F(a+1,b,c,x)-F(a,b,c,x))
\end{equation}
this can be written as
\begin{eqnarray}
<zA_{\mu}(z,x){\cal O}(0)\bar{\cal O}(y_2)> \sim \frac{1}{y_{2}^{2\Delta_1}}\frac{x_{\mu}z^{\Delta}}{x^{\Delta+1}}
F(\frac{\Delta+1}{2},\frac{\Delta-d+3}{2},\Delta-\frac{d}{2}+1,-\frac{z^2}{x^2})\nonumber \\
\label{ope3vss}
\end{eqnarray}
Finally using the identity
\begin{equation}
F(a,b,c,x)=(1-x)^{c-a-b}F(c-a,c-b,c,x)
\end{equation}
we find agreement with the $y_2 \rightarrow \infty$ limit of (\ref{mvss}).  This shows that (\ref{mvss})
has the correct asymptotic behavior for the $\mu$ components of the bulk vector.

For later use we record the 3-point function of a massive vector field strength with two boundary scalars.
It follows from (\ref{mvss}) that
\begin{eqnarray}
&& <F_{MN}(x,z) {\cal O}(y_1) \bar{\cal O}(y_2)> \nonumber\\
&& = \frac{\frac{\Delta-1}{4}}{(y_1-y_2)^{2\Delta_1}}\frac{1}{(\chi-1)^{\frac{\Delta+1}{2}}}F(\frac{\Delta+1}{2},\frac{\Delta-d+1}{2},\Delta-\frac{d}{2}+1,\frac{1}{1-\chi})\nonumber\\
&& \times \left[\partial_{M}^{x} \chi \partial_{N}^{x}\ln \frac{(x-y_1)^2 +z^2}{(x-y_2)^2 +z^2}-M\leftrightarrow N\right]
\label{MassiveFOO}
\end{eqnarray}

Now let's look at the singularity structure of these correlators, and
see if there are higher-dimension vector operators we can add to
cancel any unwanted singularities.  As in the scalar case
\cite{Kabat:2011rz} a non-zero commutator is generated between the
bulk field and the boundary operators in the region $0<\chi<1$,
corresponding to bulk spacelike separation.  Near $\chi = 1$ the
3-point function (\ref{mvss}) has an expansion for even $d$ of the
form (we present only the non-analytic terms)
\begin{equation}
\sum_{k=0}^{d/2-1} b_{k\Delta} (1-\chi)^{-\frac{d}{2}+k}+\ln (\chi-1)\sum_{k=0}^{\infty}a_{k\Delta}(1-\chi)^{k}
\end{equation}
while for odd $d$ the expansion has the form
\begin{equation}
\frac{1}{(\chi-1)^{\frac{d}{2}}}\sum_{k=0}^{\infty} a_{k\Delta}(1-\chi)^{k}
\end{equation}

For any non-conserved primary current the expansion has the same form,
just with different coefficients $a_{k\Delta}$ and $b_{k\Delta}$. For
$0<\chi<1$ this gives a non-zero commutator which is a power series in
$(1-\chi)$.  Thus if we re-define our bulk massive vector field to
include a sum of smeared non-conserved primary currents with
arbitrarily high dimension, we can cancel the commutator to whatever
order in $(1-\chi)$ we choose.  In this way we can make the bulk massive vector
obey micro-causality to an arbitrarily good approximation.

At leading order in $1/N$ we add higher dimension non-conserved currents
which are double-trace operators built from the two scalars appearing
in the 3-point function and their derivatives. For instance the
lowest-dimension primary current built in this way is
\begin{equation}
\bar{\cal O}\partial_{\mu}{\cal O}-{\cal O}\partial_{\mu}\bar{\cal O}
\label{hmvex}
\end{equation}
In the large $N$ limit this operator has dimension $2\Delta_1 +1$.  The next operator one can write is
\begin{equation}
\alpha(\nabla^{2}\bar{\cal O})\partial_{\mu}{\cal O}+\beta\bar{\cal O}\partial_{\mu}\nabla^2{\cal O}+\gamma(\partial^{\nu}\bar{\cal O})\partial_{\mu}\partial_{\nu}{\cal O}
-({\cal O}\leftrightarrow \bar{\cal O})
\label{hmvex2}
\end{equation}
This will be a primary current of dimension $2\Delta_1+3$ if
\begin{eqnarray}
\alpha & = & \frac{1}{4\Delta^{2}_1(\Delta_1+1)(d-2\Delta_1-2)} \nonumber\\
\beta & = & \frac{1}{4\Delta_1(\Delta_1+1)(d-2\Delta_1-2)}\nonumber\\
\gamma & = & \frac{1}{4\Delta_1(\Delta_1+1)}
\end{eqnarray}
A similar construction can be carried out at leading order in the
$1/N$ expansion to build primary non-conserved currents of dimension
$2\Delta_1+1+2n$ for any $n$.

\subsection{Bulk gauge fields}

Finally we turn to massless gauge fields in the bulk, where the smearing function in holographic gauge is \cite{Kabat:2012hp}
\be
z A_\mu(t,\vec{x},z) =  {1 \over {\rm vol}(S^{d-1})} \hspace{-7mm}
\int\limits_{\hspace{8mm}t'{}^2 + \vert \vec{y}^{\,\prime}\vert^2 = z^2} \hspace{-5mm} dt' d^{d-1} y' \,
j_\mu(t + t', \vec{x} + i \vec{y}^{\,\prime})
\label{MasslessGaugeSmear}
\ee
We wish to determine the higher-dimension operators which are necessary to
achieve bulk locality.  We first discuss correlators
involving the field strength, then consider the gauge field itself.

The three-point function of a bulk field strength $F_{MN}$ with two
boundary scalars can be obtained from the 3-point function of a
massive field strength with two scalars by analytically continuing
$\Delta \rightarrow d-1$.  From (\ref{MassiveFOO}) this gives
\begin{equation}
<F_{MN}(x,z){\cal O}(y_1) \bar{\cal O}(y_2)>=\frac{\frac{d-2}{4}}{(y_1-y_2)^{2\Delta_1}}\frac{1}{(\chi-1)^{\frac{d}{2}}}
\left[\partial_{M}^{x} \chi \partial_{N}^{x}\ln \frac{(x-y_1)^2 +z^2}{(x-y_2)^2 +z^2} - M \leftrightarrow N \right]
\label{GaugeFOO}
\end{equation}
This has the same singularity structure as the 3-point function for a
massive field strength, so it can be made local in the bulk in exactly
the same way, by adding appropriate smeared higher-dimension field
strengths to our definition of the bulk $F_{MN}$.  [The justification
for this analytic continuation is somewhat subtle, since the massive
vector smearing function (\ref{MassiveVectorSmearingFunction}),
(\ref{MassiveVectorSmearingFunction1}) does not smoothly go over to
the massless result (\ref{MasslessGaugeSmear}).  The first term in
(\ref{MassiveVectorSmearingFunction}), after integrating against a CFT
correlator, can be analytically continued to $\Delta = d-1$ to get the
same result one would obtain from
(\ref{MasslessGaugeSmear}).\footnote{As an example of this sort of
continuation, up to an overall normalization the result $I_1$ for a
massive vector (\ref{i1i2}) can be analytically continued to $\Delta
= d-1$ to reproduce the massless vector result (\ref{h2}) obtained
below.}  The second term in (\ref{MassiveVectorSmearingFunction}),
in the limit $\Delta = d - 1$, can be eliminated by a gauge
transformation with gauge parameter
\be
\lambda = {\Gamma(d/2) \over 2 \pi^{d/2}} \hspace{-8mm} \int\limits_{\hspace{8mm}t'{}^2 + \vert \vec{y}^{\,\prime}\vert^2 < z^2} \hspace{-5mm} dt' d^{d-1} y' \, A_z^0(t',\vec{y}^{\,\prime})
\ee
This gauge transformation also has the effect of setting $A_z$ in
(\ref{MassiveVectorSmearingFunction1}) to zero, i.e.\ it's exactly
what's needed to impose holographic gauge.]

Now consider correlators involving the bulk gauge field itself.  For
simplicity we work in the limit $y_2 \rightarrow \infty$.  In this
limit the CFT 3-point function can be obtained from (\ref{jOOlargey2})
by setting $\Delta = d - 1$.  Up to an overall constant this gives
\begin{equation}
<j_{\mu}(x) {\cal O}(0)\bar{\cal O}(y_2)> \,\,\sim \frac{1}{y_2^{2\Delta_1}} \partial_{\mu}\frac{1}{x^{d-2}}.
\end{equation}
Smearing the current into the bulk using (\ref{MasslessGaugeSmear}) gives
\begin{equation}
<zA_{\mu}(z,x){\cal O}(0)\bar{\cal O}(y_2)> \,\,\sim \frac{1}{y_2^{2\Delta_1}}\partial_{\mu}h(z,x)
\label{bgauge}
\end{equation}
where
\begin{equation}
h(z,x)=\frac{1}{{\rm vol}(S^{d-1})}\int_{t'^{2}+\vert\vec{y}\vert^{2} =z^2} d^{d-1}y dt'\frac{1}{((\vec{x}+i\vec{y})^2-(t+t')^{2})^{\frac{d-2}{2}}}
\label{h}
\end{equation}
Doing the integral in the same way as before we find
\begin{equation}
h(z,x)=\frac{z^{d-1}}{x^{d-2}}
\label{h2}
\end{equation}
Thus as $y_2 \rightarrow \infty$ we have the asymptotic behavior
\begin{eqnarray}
\nonumber&&<F_{\mu\nu}(x) {\cal O}(0) \bar{\cal O}(y_2)> \,\,\sim 0 \\
&&<F_{\mu z}(x) {\cal O}(0) \bar{\cal O}(y_2)> \,\, \sim \frac{x_{\mu}z^{d-3}}{y_2^{2\Delta_1} x^d}
\end{eqnarray}
This agrees with the ${\cal O}(1/y^{2\Delta_1})$ asymptotic behavior
of (\ref{GaugeFOO}) for $y_1 = 0$.  Since (\ref{GaugeFOO}) is AdS
covariant and has the correct asymptotic behavior, it must be the
exact result.  This is another way of seeing that the analytic
continuation we made to obtain (\ref{GaugeFOO}) is legitimate.

Now let's see what we can achieve by adding higher-dimension operators
to our definition of a bulk gauge field.  We already saw that for
correlators involving the field strength we could achieve bulk
locality by adding suitable higher-dimension currents to our
definition of a bulk vector; the massive and massless cases proceeded
in an identical manner.  However the singularity in a gauge field
correlator is not the same as for a massive vector. This can be seen
in the results above.  In the limit $y_2 \rightarrow \infty$, the
gauge field correlator (\ref{bgauge}) has singular behavior near $x =
0$, namely
\begin{equation}
<zA_{\mu}(z,x){\cal O}(0)\bar{\cal O}(y_2)> \,\,\sim \frac{1}{y_2^{2\Delta_1}} z^{d-1} {x_\mu \over x^d}
\label{Asing}
\end{equation}
In contrast, for $y_2 \rightarrow \infty$ and $x \sim 0$ the massive vector correlator (\ref{ope3vss}) behaves as
\begin{equation}
<zA_{\mu}(z,x){\cal O}(0)\bar{\cal O}(y_2)> \,\,\sim \frac{1}{y_2^{2\Delta_1}} z^{d-3} {x_\mu \over x^{d-2}}
\end{equation}
(The difference can be traced back to a cancellation between $I_1$ and
$I_2$ in (\ref{largey2AOO}).)  This means that for a gauge field one
cannot hope to cancel the boundary light-cone singularity which is present
in (\ref{Asing}) by adding massive vector fields.

This is surprising because, from the bulk point of view, we'd expect a
gauge field in the bulk to commute at spacelike separation with a
charged scalar on the boundary.\footnote{Note the step functions in
(\ref{fg}), reflecting the fact that the Wilson lines extend towards $z = 0$.}
Fortunately the requirement of having $F_{MN}$ be local
with scalars on the boundary, together with the gauge condition
$A_{z}=0$, is enough to uniquely define the bulk gauge field in terms
of smeared CFT operators.  Suppose we add a higher-dimension (hence
non-conserved) primary current $j_\mu^i$ to our definition of a bulk
gauge field, with a coefficient chosen to make $F_{MN}$ local.  Acting
with the smearing function (\ref{MassiveVectorSmearingFunction}), (\ref{MassiveVectorSmearingFunction1}) it
would seem we generate a non-zero $A_z$ in the bulk.  We can restore
holographic gauge by making a gauge transformation with parameter
\be
\lambda = \int_{0}^{z} dz' \int dx' K_{\Delta_{i}}(z',x|x')
A_{z}^{i,0}(x')
\ee
Here $\Delta_i$ is the dimension of the current
and $A_{z}^{i,0}=\frac{1}{d-\Delta_i-1}\partial^{\mu}j^{i}_{\mu}$.  So
in order to build a bulk massless vector field from boundary
operators, where the three-point function of $F_{MN}$ is local, one
has to add to the free-field definition of $A_\mu$ an infinite tower
of contributions coming from primary currents of increasing dimensions
built from ${\cal O}, {\bar{\cal O}}$ and their derivatives, of the form
\begin{equation}
\sum_{i}a_{i}\left[\frac{1}{z}\int K_{\Delta_{i}} j_{\mu}^{{i}}+\partial_{\mu}\left(\frac{1}{2(\Delta_i-d/2+1)}\int K_{\Delta_{i}+1}A_{z}^{i,0}
-\int_{0}^{z}dz' \int K_{\Delta_{i}} A_{z}^{i,0}\right)\right]
\label{GaugeCorrections}
\end{equation}
The same structure appeared in (\ref{Aeom}).  Overall the correction
(\ref{GaugeCorrections}) should make $A_\mu$ commute with boundary
scalars at spacelike separation.

\bigskip
\goodbreak
\centerline{\bf Acknowledgements}
\noindent
We are grateful to David Gross, Don Marolf, Alberto Nicolis, Joe
Polchinski, Debajyoti Sarkar, Adam Schwimmer and Steve Shenker for
valuable discussions.  The work of DK was supported by U.S.\ National
Science Foundation grants PHY-0855582 and PHY-1214410 and by PSC-CUNY
grants. The work of GL was supported in part by the Israel Science
Foundation under Grant No.\ 392/09.


\providecommand{\href}[2]{#2}\begingroup\raggedright\endgroup

\end{document}